# Multi-dimensional evaluation on a rural integrated energy system including solar, wind, biomass and geothermal energy

Ruonan Li[a], Chang Wen[a,*], Mingyu Yan[a], Congcong Wu[b], Ahmed Lotfy Elrefai[c], Xiaotong Zhang[d], Sahban Wael Saeed Alnaser[e]

[a] China-EU Institute for Clean and Renewable Energy, Huazhong University of Science and Technology, Wuhan 430074, China.

[b] Key Laboratory for the Green Preparation and Application of Functional Materials, Hubei Key Laboratory of Polymer Materials, School of New Energy and Electrical Engineering, Hubei University, Wuhan, China

[c] Faculty of Engineering, Egypt Japan University of Science and Technology (EJUST), Egypt.

[d] State Grid Liaoning Electric Power Research Institute Co., Ltd., Shenyang, China.

[E] Department of Electrical Engineering, University of Jordan, Jordan.

*CORRESPONDING AUTHOR:

**Chang Wen.** Associate Professor. Email: wenchang@hust.edu.cn.

Mailing address: Department of New Energy Science and Engineering, School of Energy and Power Engineering, Huazhong University of Science and Technology, 1037 Luoyu Road, Wuhan 430074, China.

**Abstract:** This study focuses on the novel municipal-scale rural integrated energy system (RIES), which encompasses energy supply and application. By constructing a seven-dimensional evaluation system including energy efficiency, energy supply, low-carbon sustainability, environmental impact, energy economy, social benefits, and integrated energy system development, this research combines the improved analytic hierarchy process (IAHP) and entropy weight method (EWM) by sum of squares of deviations to balance expert experience and data objectivity. Furthermore, the cloud model is introduced to handle the fuzziness and randomness in the evaluation. This method can quantify the differences in system performance before and after the planning implementation. The results indicate that after planning, the comprehensive score has increased from 83.12 to 87.55, the entropy value has decreased from 6.931 to 5.336, indicating enhanced system stability. The hyper-entropy has dropped from 3.08 to 2.278, reflecting a reduction in uncertainty. The research findings provide a scientific basis for the planning optimization, policy-making, and sustainable development of rural integrated energy systems, possessing both theoretical innovation and practical guiding value.

# 1. Introduction

The coordinated development of rural energy systems with agriculture, industry, rural tourism, and other industries is a crucial aspect of rural revitalization for China. The current disconnect between energy supply and demand, which leads to substantial resource waste, necessitates urgent optimization of distributed energy system layouts, the establishment of multi-energy complementary mechanisms, and the implementation of intelligent regulation strategies to effectively manage energy fluctuations.[1]. By analyzing technical viability, economic feasibility, and environmental benefits, multi-dimensional evaluation multi-dimensional evaluation as a key tool for coordinating multi-energy complementarity and green low-carbon development, not only identifies system strengths and weaknesses, but also clarifies optimization directions to enhance overall efficiency. Additionally, it scientifically assesses carbon reduction effects, ensuring compliance with China's national goals of "carbon peaking by 2030 and carbon neutrality by 2060", thereby offering scientific foundations for policy formulation and resource allocation to facilitate green transformation and high-quality rural energy development.

In current research on novel rural integrated energy system evaluation, inconsistencies persist in conceptual definitions, calculation methods, and interpretations of identical indicators, creating challenges in consolidating and comparing assessment results. Existing quantification methods inadequately measure indicators involving complex social, environmental, and system operational characteristics and their impacts, compromising the scientific validity and reliability of evaluations. Regarding indicator systems, the concerned researchers [2–15] discussed evaluation metrics for integrated energy systems across energy utilization, economics, environment, system performance, social benefits, and smartization. For weight determination, the analytic hierarchy process (AHP) effectively combines qualitative and quantitative analyses through its hierarchical structure to derive subjective weights [16–18]. The entropy weight method (EWM) determines the objective weight based on the degree of data variation, thereby reflecting the inherent significance of the indicators [19–22]. However, single weighting methods fail to balance subjective expertise and objective data. Concerned researchers [23–30] explored combined weighting approaches (maximum deviation, game theory, deviation square sum) to investigate synergistic optimization mechanisms in multi-attribute decision-making, demonstrating enhanced robustness and stability of decision models. Compared to the fuzzy evaluation method employed in conventional assessments which only integrate qualitative and quantitative indicators [31], the cloud model more precisely quantifies uncertainty by defining risk ranges (via entropy) and reflecting result stability (via hyper-entropy) [32,33]. Although widely applied in risk assessment and performance evaluation, its use in integrated energy system assessments remains limited [34,35]. Nie et al. [36] established a 7-indicator system (energy, economy, environment) using game theory-combined AHP-EWM weights and cloud modeling to optimize industrial park energy systems. Validation via game theory-enhanced Grey Relational Analysis (GRA) and TOPSIS methods confirmed model robustness, showing solution consistency and priority stability with cloud model results.

Addressing existing challenges in rural integrated energy system evaluation, this study capitalizes on the cloud model's potential for handling uncertainties in energy efficiency applications. We develop a novel evaluation framework with seven dimensions: energy efficiency, energy supply, low-carbon sustainability, environmental impact, energy economy, social benefits, and integrated energy system development. Each selected indicator is explicitly defined and quantitatively formulated. Methodologically, the improved analytic hierarchy process (IAHP) calculates normalized eigenvectors to obtain subjective weight coefficients, while EWM computes objective weight coefficients through information entropy. The combined weighting method based on the sum of squares of deviations integrates these weights.

Evaluation grades for rural integrated energy systems are defined, and corresponding cloud model parameters are determined. Based on indicator data and combined weights, the indicator cloud parameters and integrated cloud model parameters are calculated. The system's integrated cloud model parameters are compared with standard cloud model parameters. Following the maximum similarity principle, the hierarchical evaluation grades and comprehensive assessment results are determined. Subsequently, this study empirically validates the novel rural energy system in a rural energy revolution pilot county, during the pre- and post-2025 planning stages, verifying the feasibility of the assessment methodology while evaluating the effectiveness of the energy planning strategy.

## 2. Theory overview

### 2.1 Construction of novel evaluation index for rural integrated energy system

The comprehensive assessment of integrated energy system development should consider four key dimensions — generation, grid, load, and storage [37] — along with multi-criteria indicators including efficiency, environmental, economic, and social metrics, as shown in the Fig. 1.

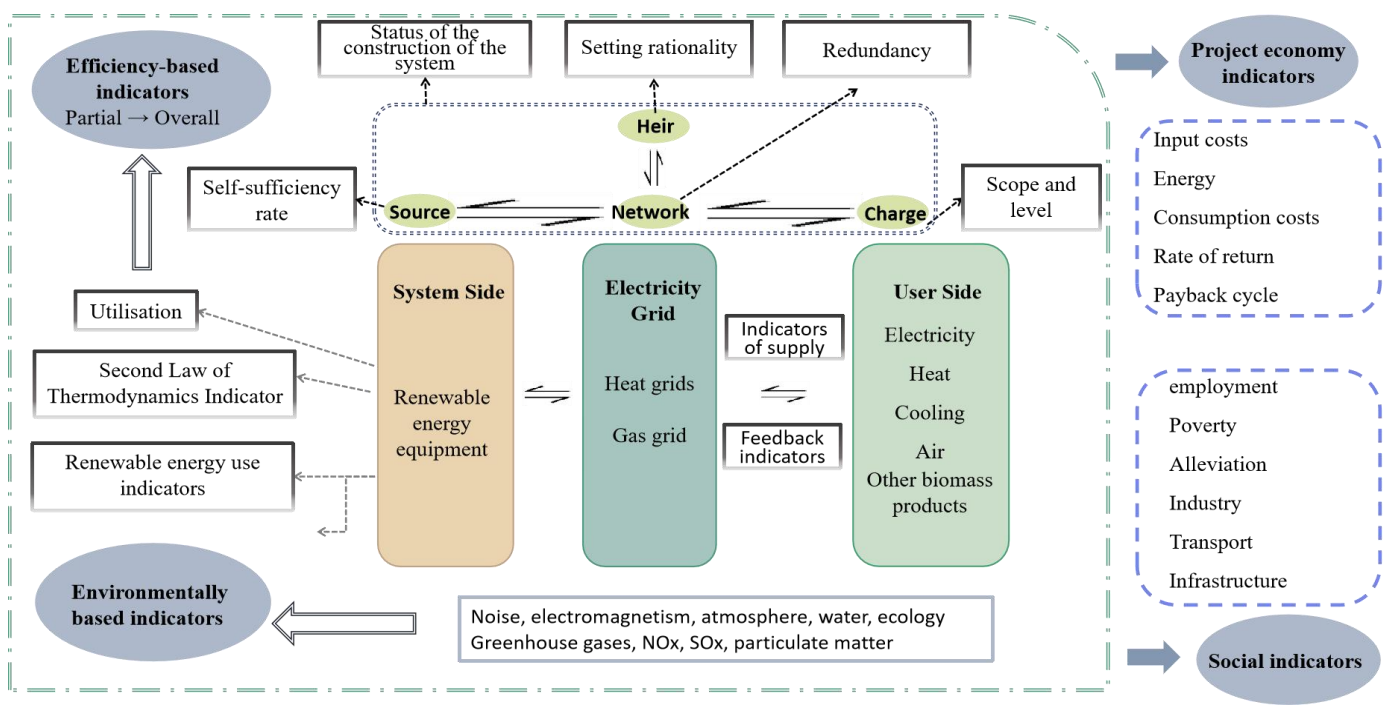

**Fig. 1.** Schematic diagram of multidimensional indicator relationships in integrated energy systems.

The generation focus on the energy self-sufficiency rate, which reflects the system's capability to meet energy demands through its own resources. A higher self-sufficiency rate directly enhances the stability and autonomy of the energy system. The grid centers on the energy redundancy level, where sufficient redundancy ensures supply reliability, guaranteeing stable operation of the energy network during emergencies or peak demand periods to avoid service interruptions. The load emphasizes the load range and magnitude, where understanding temporal and spatial demand characteristics facilitates rational energy supply planning and improves utilization efficiency. The storage depends on the rational configuration of energy storage equipment, as appropriate storage deployment effectively balances temporal supply-demand mismatches, enhancing system flexibility and stability [38].

Beyond the four core dimensions of generation, grid, load, and storage, comprehensive evaluation of energy systems necessitates critical attention to diverse indicator types. Efficiency metrics are paramount for system performance: improving the utilization rates of energy and equipment to reduce waste and enhances efficiency, while exergy efficiency analysis grounded in the Second Law of Thermodynamics quantifies irreversible losses during energy conversion, offering theoretical insights for system optimization [39]. Concurrently, renewable energy utilization metrics serve as vital benchmarks for sustainability and system greening. Environmental metrics require rigorous assessment of operational impacts, including noise, electromagnetic emissions, air pollution, water contamination, and ecological disturbances caused by energy infrastructure. During the energy utilization process, the emissions of greenhouse gases (GHG), nitrogen oxides ($NO_x$), sulfur oxides ($SO_x$), and particulate matter (PM) have a direct impact on environmental quality and ecological balance. To this end, the Life Cycle Assessment (LCA) method [40] can be employed to comprehensively quantify the pollutant emissions throughout the entire life cycle of energy systems, from production and transportation to usage and disposal. Supply reliability metrics evaluate the performance of power grids, thermal networks, and gas distribution systems in delivering energy with consistent quality and stability. User feedback on energy usage provides critical insights for optimizing energy networks, driving iterative improvements to better align with demand patterns.

For the entire energy project, economic and social indicators are equally significant. Economic indicators primarily encompass the total life cycle cost of the project, energy consumption costs, rate of return, payback period, and more. These indicators are directly related to the project's economic benefits and sustainability. Social metrics evaluate local impacts through employment generation, poverty alleviation, industrial development, transportation upgrades, and infrastructure enhancement. An exemplary energy project must deliver both economic returns and tangible societal contributions.

Integrating these considerations, the constructed rural integrated energy evaluation system is shown in the Fig. 2 below. The first-level index is the objective Layer, which represents comprehensive evaluation outcomes. The second-level indicators are considered from seven aspects: energy efficiency (C1), energy supply (C2), low-carbon sustainability (C3), environmental impact (C4), energy economy (C5), social benefits (C6), and integrated energy system development (C7).

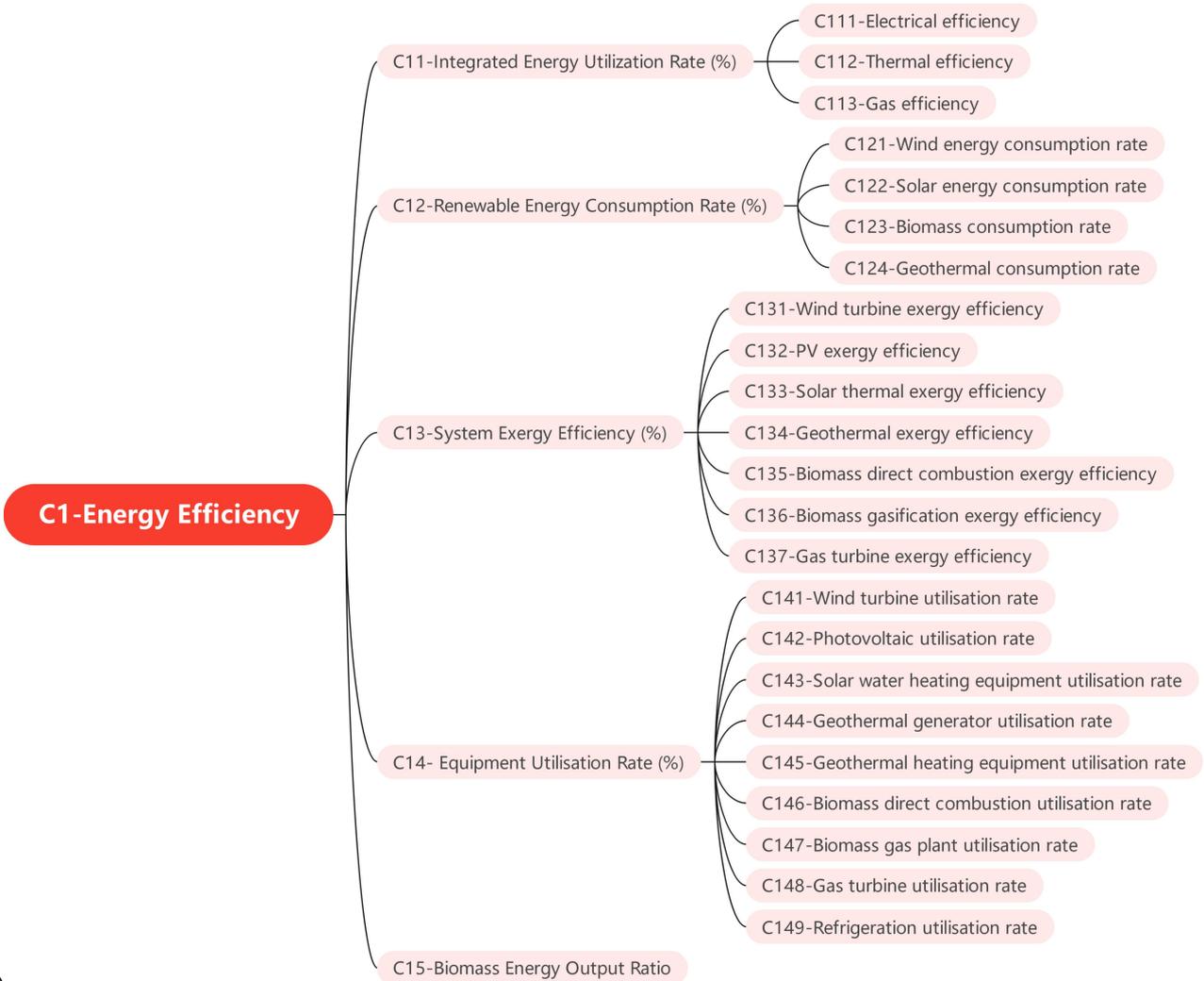

(a)

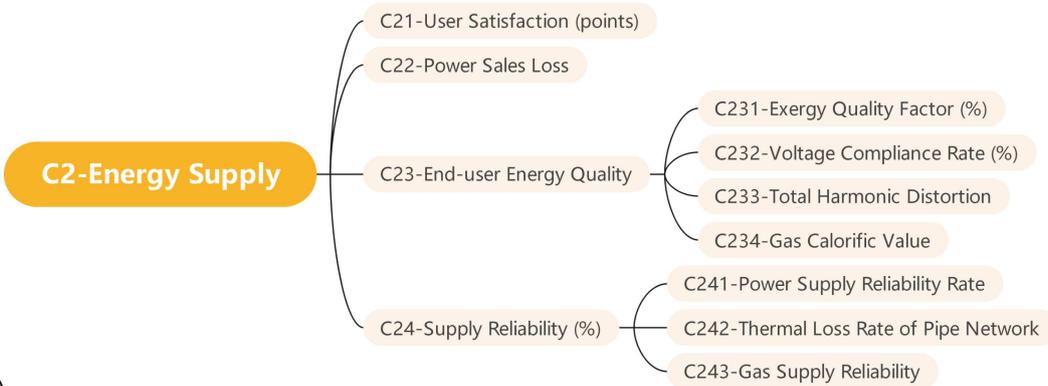

(b)

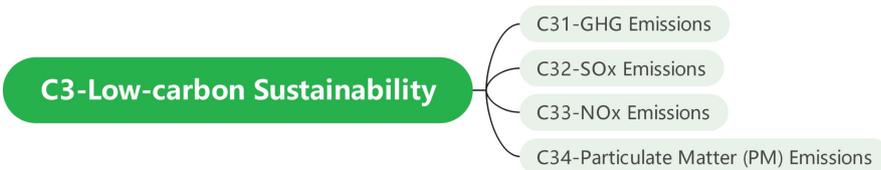

(c)

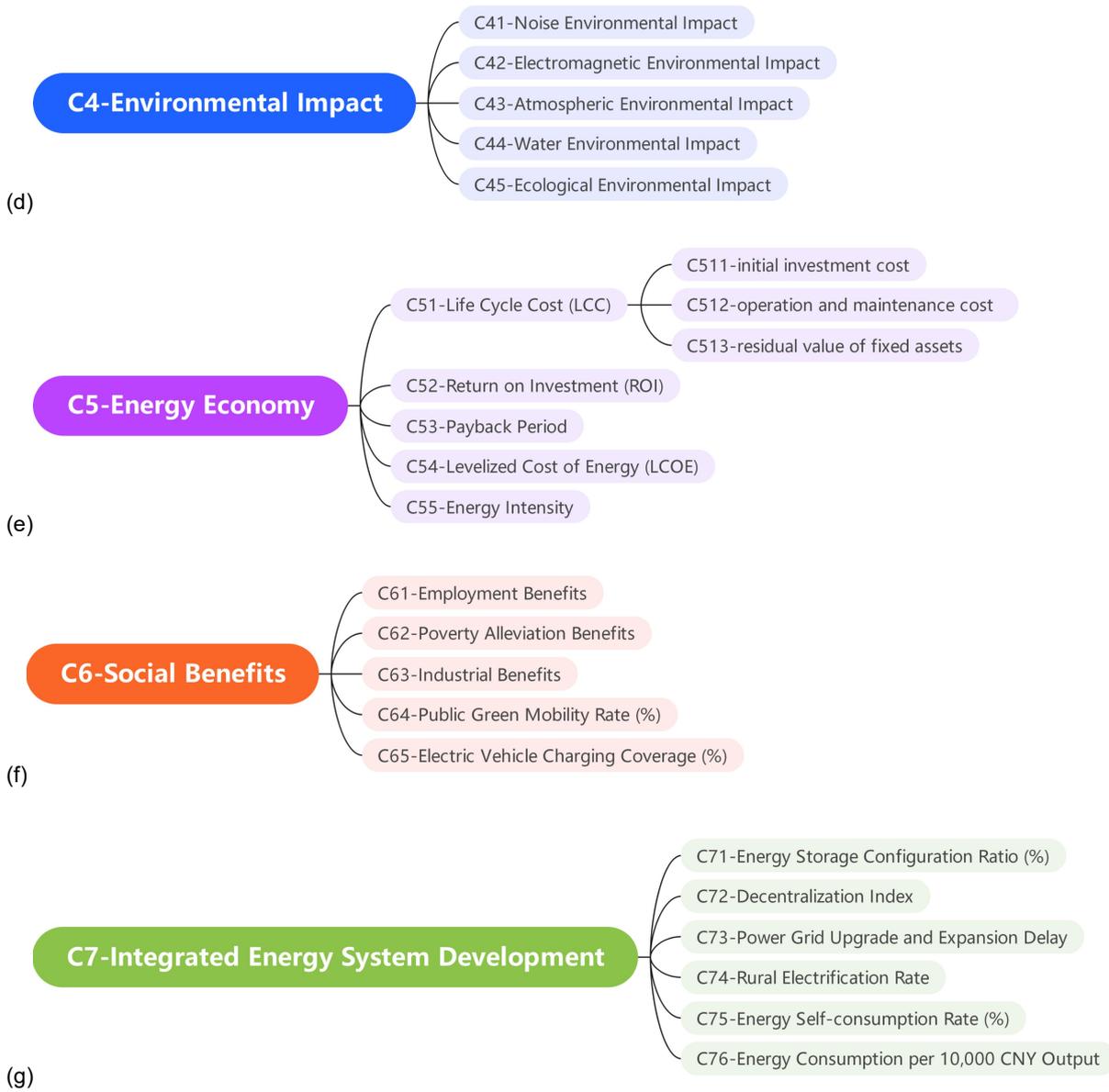

**Fig. 2.** The developed evaluation index for rural integrated energy system.

Energy efficiency (C1) encompasses indicators including Integrated Energy Utilization Rate, System Exergy Efficiency, Renewable Energy Accommodation Rate, Equipment Utilization Rate, and Biomass Energy Output Ratio, which collectively assess energy utilization efficiency and renewable energy integration at the tertiary indicator level[2,11,24,39,41]. Energy supply (C2) incorporates indicators such as User Satisfaction Index, Power Sales Loss, End-user Energy Quality, and Supply Reliability, reflecting user experience, service quality, and operational dependability during energy delivery [3,4,16]. Low-carbon sustainability (C3) evaluates environmental impacts through GHG Emissions, $SO_x$ Emissions, $NO_x$

Emissions, and Particulate Matter (PM) Emissions, quantifying the ecological footprint of rural integrated energy systems[7,13]. Environmental Impact (C4) encompasses indicators including Noise Environmental Impact, Electromagnetic Environmental Impact, Atmospheric Environmental Impact, Water Environmental Impact, and Ecological Environmental Impact, systematically categorizing different dimensions of environmental consequences[9,38]. Energy economy (C5) covers Life Cycle Cost, Return on Investment, Payback Period, Levelized Cost of Energy, and Energy Intensity to appraise economic viability and cost-effectiveness[5,10,17]. Social benefits (C6) addresses Employment Generation Rate, Poverty Alleviation Index, Industrial Value-added Rate, Public Green Mobility Rate, and EV Charging Coverage, demonstrating societal contributions[1,6]. In terms of Integrated Energy System Development (C7), there are indicators such as Energy Storage Configuration Ratio, Decentralization Index, Power Grid Upgrade and Expansion Delay, Rural Electrification Rate, Energy Self-consumption Rate, and Energy Consumption per 10,000 CNY Output, characterizing system deployment and operational dynamics[8,33,42]. Detailed definitions of all indicators (including tertiary-level sub-indicators) and their corresponding mathematical formulations can be found in the Section A1 of the Supplementary Materials.

Based on these indicators, a multi-level evaluation framework is established, comprising the Objective Layer (overarching goals), Criterion Layer (key dimensions), and Indicator Layer (specific metrics). These hierarchically linked tiers form a cohesive system for multi-dimensional performance assessment of new rural integrated energy systems.

## 2.2 Evaluation of the novel rural integrated energy system

This paper combines three methods — analytic hierarchy process (AHP), entropy weight method (EWM), and deviation square sum combined weighting — to rationally determine comprehensive subjective-objective weights for all indicators. A cloud model-based evaluation framework is constructed for rural energy systems, with matrix operations implemented using MATLAB.

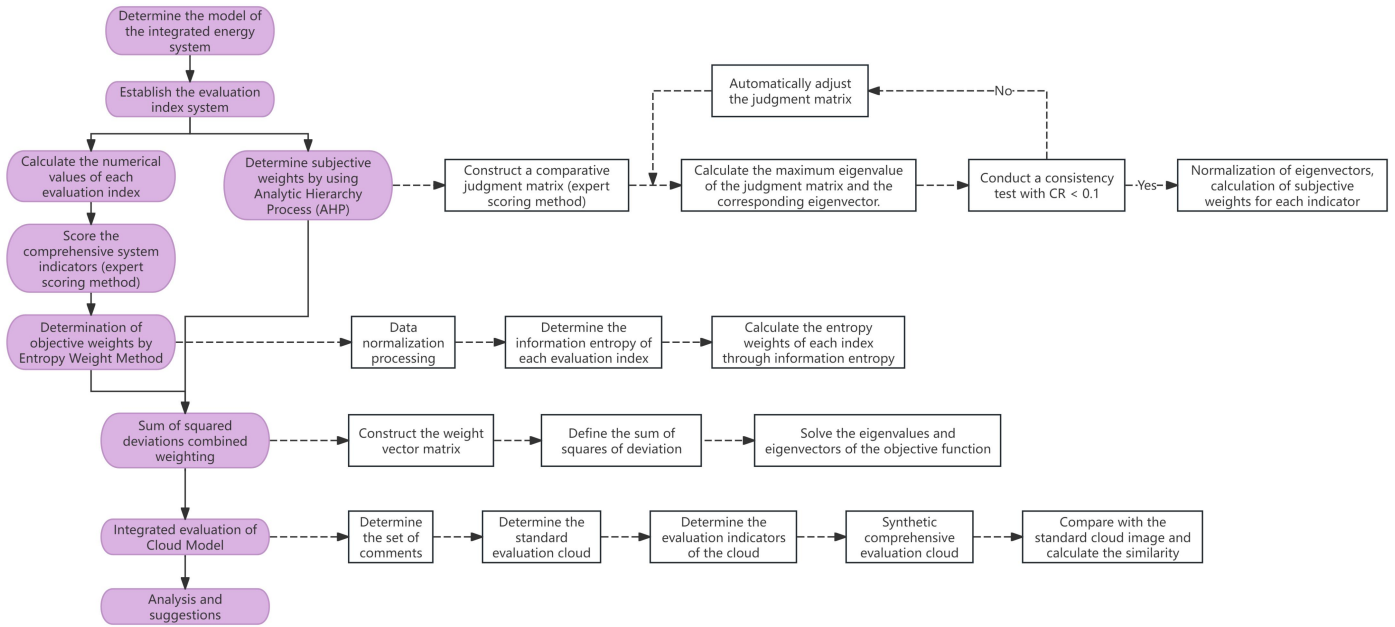

**Fig. 3.** Evaluation process of the novel rural integrated energy system.

**2.2.1 Improved analytic hierarchy process (IAHP) to determine the subjective weights**

AHP quantifies index weights by constructing hierarchical structure and judgment matrix in novel integrated energy system evaluation. However, the traditional AHP needs to manually modify the judgment matrix, which has some problems such as low efficiency, easy to be disturbed by subjective factors, and difficult to ensure consistency. The introduction of automatic correction mechanism can significantly improve decision-making efficiency, reduce subjective bias, and facilitate large-scale application [43]. The calculation methods and formulas of traditional AHP are included in Section B1 of the Supplementary Materials.

During the construction of a judgment matrix, if experts assign contradictory values to pairwise comparisons of indicators, the matrix will exhibit inconsistency. The consistency ratio (CR) is a key metric used to test the logical consistency of the judgment matrix. By mathematically detecting the degree of such inconsistency, this method ensures the reliability of weight calculation. In establishing the (AHP), it was asserted that CR < 0.1 indicates acceptable consistency of the judgment matrix, CR > 0.1 necessitates revisions to the matrix to resolve logical inconsistencies [43]. Automatic correction process is as follow.

First, the scale transformation of the judgment matrix $R = (r_{ij})_{n \times n}$ is carried out, and the mapping relationship is shown in the following Table 1.

Table 1 The mapping relationship between $r_{ij}$ and $r^*_{ij}$.

| $r_{ij}$ | $r^*_{ij}$ | $r_{ij}$ | $r^*_{ij}$ |
| --- | --- | --- | --- |
| 1/9 | 0.1 | 2 | 0.55 |
| 1/8 | 0.15 | 3 | 0.6 |
| 1/7 | 0.2 | 4 | 0.65 |
| 1/6 | 0.25 | 5 | 0.7 |
| 1/5 | 0.3 | 6 | 0.75 |
| 1/4 | 0.35 | 7 | 0.8 |
| 1/3 | 0.4 | 8 | 0.85 |
| 1/2 | 0.45 | 9 | 0.9 |
| 1 | 0.5 | | |

For j>i+1,

$$\bar{r}^*_{ij} = \frac{\sqrt[j-i-1]{\prod_{t=i+1}^{j-1} r^*_{it} r^*_{tj}}}{\sqrt[j-i-1]{\prod_{t=i+1}^{j-1} r^*_{it} r^*_{tj}} + \sqrt[j-i-1]{\prod_{t=i+1}^{j-1} (1-r^*_{it})(1-r^*_{tj})}} \tag{1}$$

For j≤i+1, $\bar{r}^*_{ij} = r^*_{ij}$

Calculate the distance metric.

$$(R^*, \bar{R}^*) = \|R^* - \bar{R}^*\| = \sqrt{\sum_{i=1}^{m} \sum_{j=1}^{n} |r^*_{ij} - \bar{r}^*_{ij}|^2} \tag{2}$$

Generally, the consistency threshold τ is set to 0.1, which is a reference to the common judgment standard for consistency ratio in traditional AHP. If $d(R^*, \overline{R}^*) < 0.1$, the consistency indicated $\overline{R}^*$ is acceptable, If $d(R^*, \overline{R}^*) > 0.1$, it is inconsistent and needs to be repaired.

The repair algorithm is:

$$\tilde{r}^*_{ij} = \frac{(r^*_{ij})^{1-\sigma}(\bar{r}^*_{ij})^{\sigma}}{(r^*_{ij})^{1-\sigma}(\bar{r}^*_{ij})^{\sigma}+(1-r^*_{ij})^{1-\sigma}(1-\bar{r}^*_{ij})^{\sigma}} \quad (3)$$

σ is the control parameter determined by the decision maker. The smaller the value $\sigma$ is, the closer $\widetilde{R}^*$ is to $R^*$. The parameter settings, σ = 0.8 here, align with the criteria established by Xu et al. [43].

Calculate the distance measure $d(\overline{R}^*, \widetilde{R}^*)$ again, if $d(\overline{R}^*, \widetilde{R}^*) < 0.1$, then the consistency of the statement $\widetilde{R}^*$ is acceptable; If $d(\overline{R}^*, \widetilde{R}^*) > 0.1$, it is inconsistent and needs to be repaired again. Repeat until the requirements are met.

After meeting the requirements, the reverse transformation of the above scale is carried out, from 0.1~0.9 scale to 1/9 ~ 9 scale, and the revised judgment matrix R is output.

At this time, the consistency test is carried out CR < 0.1, and the judgment matrix $R = (r_{ij})_{n \times n}$ meets the consistency conditions.

**2.2.2 Entropy weight method for objective weights**

EWM is a method for comprehensive evaluation of multi-indicators and multi-objects. We take multiple evaluation results as objects, and use EWM to objectively assign weights to indicators and reflect the influence of different indicators on evaluation results. The specific calculation process and formula of EWM are shown in Section B2 of the Supplementary Materials.

**2.2.3 Deviation square sum combined weighting**

The deviation square sum combined weighting method identifies an optimal composite weighting coefficient that maximizes the total deviation square sum of all evaluation indicators under the combined weights, thereby deliberately amplifying disparities in weight allocation among indicators. This prevents equalization of weights caused by increases in indicator scale and ensures that evaluation results exhibit stronger distinguishing validity and informational completeness in a statistical sense. In integrating subjective and objective weights, the method preserves the capability of subjective weighting to differentially prioritize core indicators—ensuring consistency between the evaluation system and decision-making objectives—while introducing data-driven characteristics of objective weighting to suppress distortion risks arising from expert knowledge blind spots or group preferences [44]. The specific calculation process and formula of deviation square sum combined weighting are shown in Section B3 of the Supplementary Materials.

## 2.3 Evaluation model of novel rural integrated energy system based on cloud model

### 2.3.1 Basic theory of fuzzy comprehensive evaluation and cloud model

In the research of multi-criteria decision-making and comprehensive evaluation, the fuzzy comprehensive evaluation (FCE) method has been widely applied due to its ability to handle fuzzy information. Its core advantage lies in using the fuzzy transformation technology to convert the evaluation results of discrete factors into comprehensive evaluation values, effectively solving the problem that traditional evaluation methods have insufficient depiction of fuzziness. However, this method still has several limitations. Firstly, there is no unified standard for determining the membership function, which often relies on expert experience, making the evaluation results vulnerable to subjective factors. Secondly, when dealing with high-order fuzziness and uncertainty, its mathematical framework has difficulty in effectively capturing the random factors in the evaluation process.

To resolve these issues, the cloud model proposed by academician Li Deyi establishes a bidirectional cognitive transformation mechanism between qualitative concepts and quantitative data, integrating fuzziness and randomness [45]. By analyzing dual uncertainties in linguistic variables through probabilistic measure theory, this model reveals intrinsic connections between fuzziness and stochasticity, significantly enhancing decision-making efficacy for linguistic data. Compared to conventional fuzzy methods, the cloud model demonstrates unique advantages in evaluation objectivity and systemic uncertainty characterization [46]. The definition and computational process of cloud model are shown in Section C of the Supplementary Materials.

In the cloud model, the digital characteristics of the cloud are often represented by three values (as shown in the Fig. 4): expected value (Ex), entropy (En) and hyper-entropy (He), which combine fuzziness and randomness together to form a qualitative and quantitative mapping. It provides a convenient tool for the processing of qualitative and quantitative index system [41].

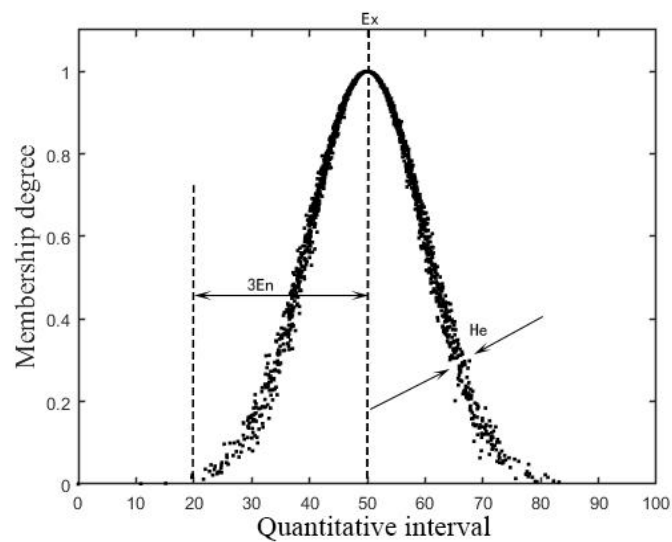

**Fig. 4.** Digital characteristics of the qualitative concept A cloud.

Ex is expected to be the point that best represents the qualitative concept A in the number domain space. In essence, it is the typical sample point for this qualitative concept at the quantitative level. Among many possible samples, Ex like the core representative of qualitative concept A, embodies the key feature of the concept, which is the exact quantitative mapping of the qualitative concept in the number domain space.

En is an important index to reflect the uncertainty of qualitative concept A, and its uncertainty is reflected in the aspects of fuzzy degree, randomness, fuzziness and randomness correlation and concept granularity. In terms of vagueness, En measures the size of the range of cloud droplet groups that can be accepted by the linguistic value A in the number domain space. In terms of randomness, En also represents the probability that the cloud droplet group in the number domain space can represent the language value. It reflects the randomness of the occurrence of a cloud droplet representing a qualitative concept.

He is an uncertainty measure of En and can be understood as "the entropy of entropy". It reflects the cohesion of the uncertainty of all points representing the linguistic value in the numerical domain space, that is the cohesion degree of the cloud drops. The larger the He is, the greater the dispersion degree of the cloud drops will be, which means that the randomness of the membership degree is higher and the "thickness" of the cloud is also greater.

## 3. Results and discussion

### 3.1 Case selection and data collection

Based on the investigation of the construction and operation status of the renewable energy system, this study systematically sorted out the rural renewable energy development model and energy supply system in combination with the 2025 planning goals in the "Implementation Plan for the Construction of the Pilot County of Rural Energy Revolution". By integrating the energy demand of farmers' living and agricultural production, the model framework of the novel rural integrated energy system is constructed, as shown in Fig. 5.

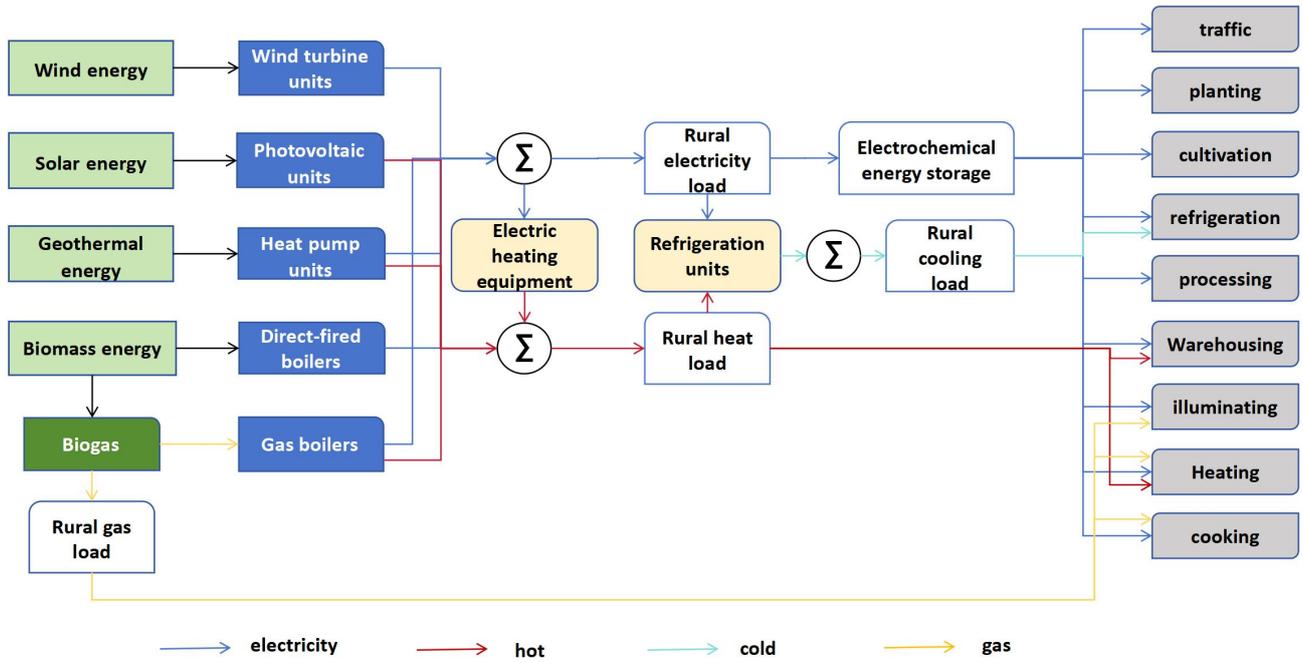

**Fig. 5.** Novel rural integrated energy system model framework.

On the energy supply side, the system integrates multiple energy conversion technologies: wind turbines achieve wind-energy conversion, photovoltaic arrays complete solar-energy conversion, and geothermal resources generate electricity and heat simultaneously through cogeneration units. Biomass energy uses a dual path utilization mode -- direct combustion generator sets directly convert biomass into electricity, or through anaerobic fermentation process to produce biogas, the latter can be directly supplied to users through the pipe network, but also through the gas cogeneration unit to achieve cascade utilization [48].

In the energy conversion link, the system has built a multi-energy complementary coupling network: electric energy can be converted into heat energy through the electric heating device, or drive the compression refrigeration unit to produce cold energy. Waste heat resources can be converted into cold energy by absorption refrigeration system, forming a trinity of "heat-electric-cold" energy conversion system. This multi-energy collaborative architecture significantly improves the comprehensive energy utilization efficiency, and provides a technical paradigm for the low-carbon transformation of rural energy systems.

Based on relevant government planning documents, corporate collaboration projects, relevant industry statistics and field research cases, we have collected multi-dimensional data — including technical parameters, operational data, and economic indicators of the rural integrated energy system. That provides robust support for the construction and validation of analytical models. After the implementation of Tianmen City's integrated energy system plan, renewable energy has become the core component of regional energy consumption. Wind and solar power installed capacity has achieved leapfrog growth with large-scale supporting energy storage facilities. Distributed wind power development and agro-photovoltaic integration models have extended the industrial chain for new energy equipment manufacturing and intelligent operation/maintenance, driving significant growth in the local economy (2.3% GDP increase) and creating 12,000 jobs. Additionally, the city has established coordinated optimization across the "generation-grid-load-storage" chain, forming a new energy system driven by dual forces of technological iteration and industrial-economic collaboration.

## 3.2 Weight calculation and result analysis

As can be seen from the statistical results in the following Fig. 6, the subjective weight can significantly reflect the difference in the importance of each indicator and fully consider the complex relationship among indicators. Objective weights coordinate the importance of each indicator according to the data itself. The deviation square sum combined weighting not only retains the difference reflected in the subjective weighting, but also avoids the weight deviation caused by the expert's knowledge limitation or personal bias with the help of the objective weighting.

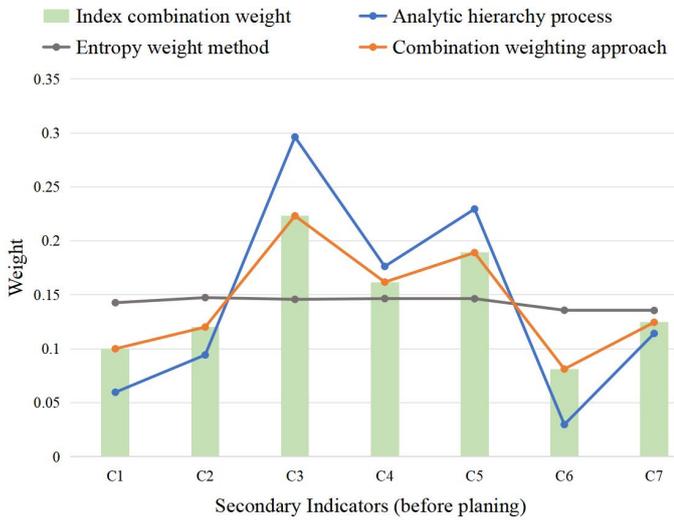

(a)

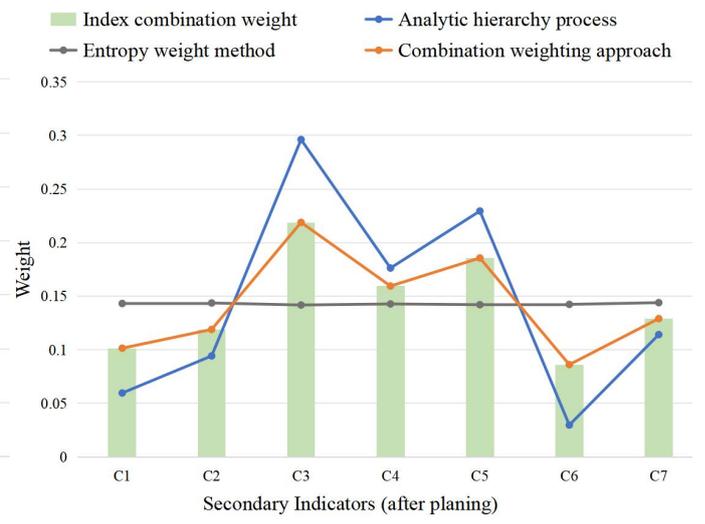

(b)

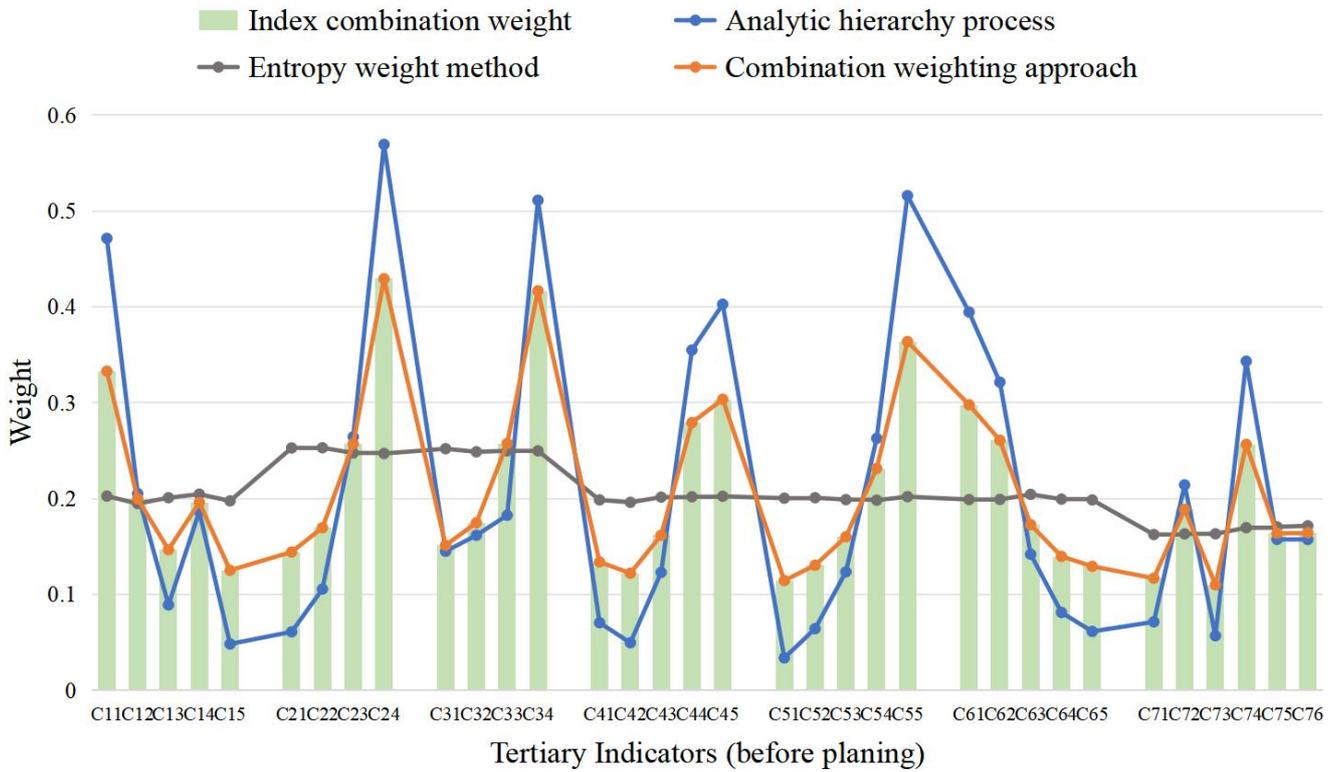

(c)

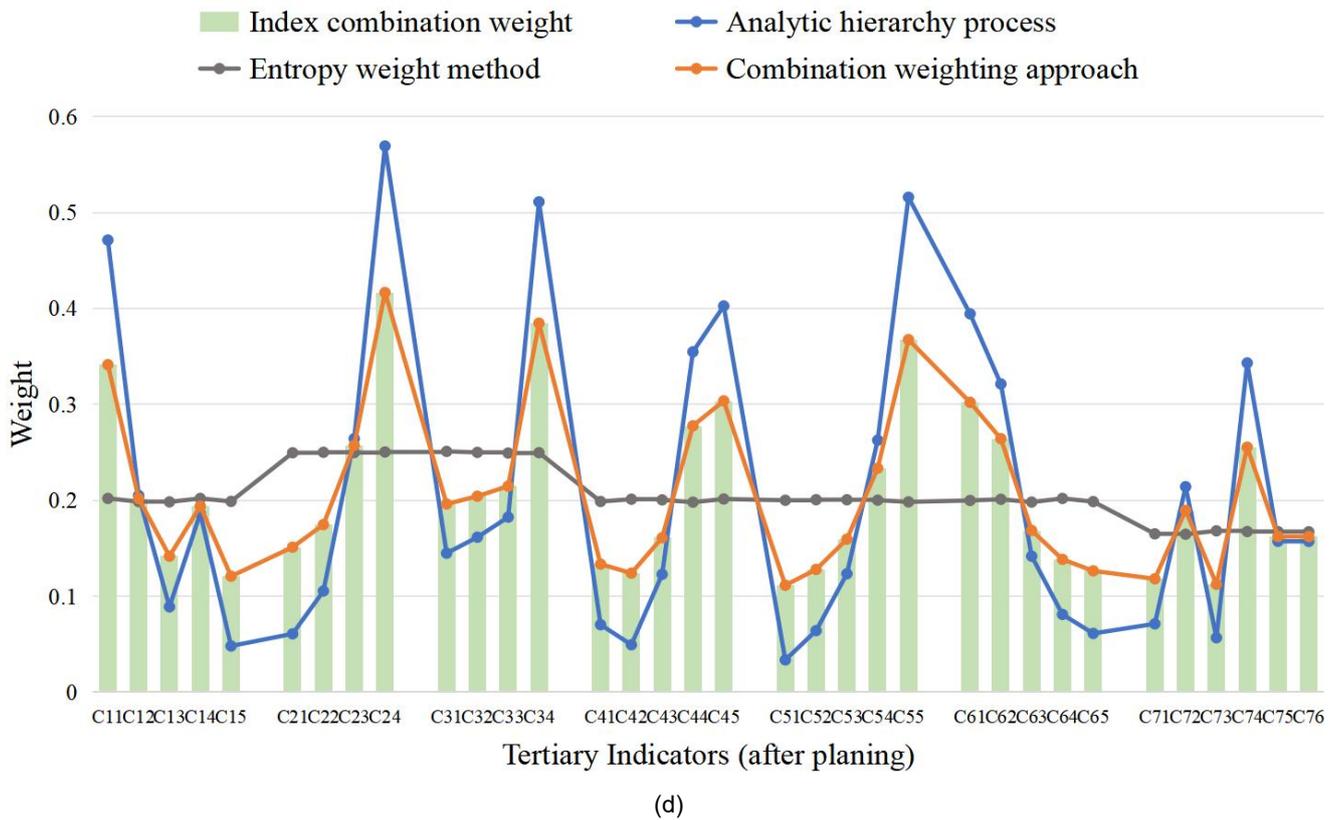

**Fig. 6.** The weight comparison diagrams of indicators before and after planning under multiple weighting methods: (a) secondary indicators weight comparison before planning, (b) secondary indicators weight comparison after planning, (c) tertiary indicators weight comparison before planning, (d) tertiary indicators weight comparison after planning.

By comparing Fig. 6(a) with Fig. 6(b) and Fig. 6(c) with Fig. 6(d), respectively, it can be found that the weights of secondary and tertiary indicators before and after the planning are basically consistent, with only slight differences, indicating the limited impact on the evaluation results before and after the planning. Among the secondary indicators, C3 and C5 have the highest weight, which indicates that in the current index system, low-carbon index and economic index are of great importance; While the low weight of C6 indicates that the social index plays a relatively small role in the system. Among the tertiary indicators, the Integrated Energy Utilization Rate (C11) is relatively important among the Energy Efficiency (C1), the Supply Reliability (C24) is relatively important among the Energy Supply (C2), the PM Emissions (C34) is relatively important among the Low-carbon Environmental Protection (C3), the Water Environment Impact (C44) and Ecological Environment Impact (C45) are relatively important among the Environmental Impact (C4), the Energy Intensity (C55) is relatively important among the Energy Economy (C5), the Employment

Benefits (C61) and Poverty Alleviation Benefits (C62) are relatively important among the Social Benefits (C6), and the Rural Electrification Level (C74) is relatively important among the Integrated Energy System Construction (C7). The results show that the important indicators are consistent with those most frequently used by relevant researchers in the indicator system survey.

### 3.3 Evaluation results and analysis based on cloud model

By comparing the evaluation results of cloud model with those of fuzzy evaluation method, it can be seen that before planning, the score of fuzzy evaluation is 84.048, and the score of cloud model evaluation is 83.118. After planning, the score of fuzzy evaluation was 88.05, and the score of cloud model evaluation was 87.547. Both methods show that the score after planning is higher than before planning, and the change trend is consistent, which indicates that the cloud model evaluation results are accurate to a certain extent, and can effectively reflect the state changes of the integrated energy system before and after planning.

Through the fuzzy evaluation results, we can only get a slight improvement in the overall score of Tianmen comprehensive energy system after the planning. However, more multidimensional analysis can be carried out according to Fig. 7. Before the planning, the Ex is 83.118, and after the planning, it is increased to 87.547. This increase indicates that the overall performance or status of the comprehensive energy system in Tianmen City has been significantly improved after the planning, and the planning measures have effectively promoted the system to develop in a better direction. The En before planning is 6.931, and after planning it drops to 5.336. The state of the integrated energy system is more clearly defined and the system state is more stable, indicating that the planning makes the operation and evaluation of the system more standardized and predictable. The He is 3.08 before planning, and decreases to 2.278 after planning, indicating that the cohesion of uncertainty of various factors in the integrated energy system after planning is weakened, the relationship between various elements within the system is more stable and orderly, the system runs more smoothly, and the fluctuations and risks brought by uncertainty are reduced.

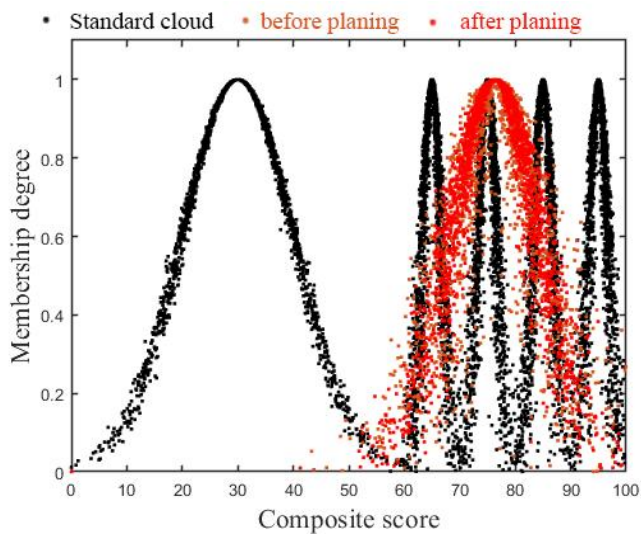

**Fig. 7.** Tianmen City's integrated energy system before and after the planning. Comprehensive cloud diagram.

The further analysis of the cloud map of the two indexes before and after the planning shows that the core indexes of the integrated energy system of Tianmen City show an overall optimization trend after the planning. Ex of all indicators increased significantly, and Ex of Energy Efficiency increased from 82.08 to 87.94, indicating that the integration of multiple energy conversion technology significantly improved the stability and efficiency of energy supply. However, Ex of wind turbines and photovoltaic arrays increased and En decreased. It shows that the fluctuation of wind-solar output is effectively controlled through the optimization of weather prediction model and energy storage configuration (such as the smooth output of lithium batteries). The Ex of Social Benefit jumped from 76.52 to 87.28, an increase of 14.1%, reflecting the enhanced response ability of the energy system to social demand, which may be closely related to the improvement of power supply stability in rural areas and the increase of employment opportunities in the new energy industry. It is worth noting that the Ex of Integrated Energy System Development increased mostly

(76.34→88.06, +15.3%), highlighting the effectiveness of the integration of multi-energy complementary networks and infrastructure.

With the exception of a slight increase in the En of C1, the En of other indicators showed a downward trend. The En of C4 decreased from 8.038 to 5.641, a decrease of 29.8%, indicating that the system's ability to predict and control environmental risks such as pollutant discharge and ecological disturbance was significantly enhanced; The En of C5 decreased from 5.924 to 4.666, indicating that the volatility of the relationship between energy and economy was reduced and the stability of income was improved. The small increase in En for C1 (4.797→5.295) may be due to the uncertainty caused by the commissioning of new technologies (such as waste heat refrigeration units) in the early planning phase, but the decrease in He (3.435→2.412) indicates stability in the long run.

As shown in the Fig. 8, in addition to the increase in the He value of C7, the He of other indicators has decreased, and the He of C3 has decreased from 3.489 to 1.867, a decrease of 46.5%, indicating that after the proportion of renewable energy has increased (such as the large-scale application of wind power and photovoltaic), C32 have been greatly reduced, and the effect of policy implementation is more concentrated. The rise in He value of C7 is essentially the reflection of the uneven technical maturity of its internal subsystems and the initial volatility of complex system integration. The mutual coordination of distributed energy and the high He value of C74 the optimization effect of energy storage, power grid upgrading and other fields. In the long run, with the completion of technical debugging and the improvement of data governance, the comprehensive He value will gradually decline, and the system stability and synergy will be significantly improved.

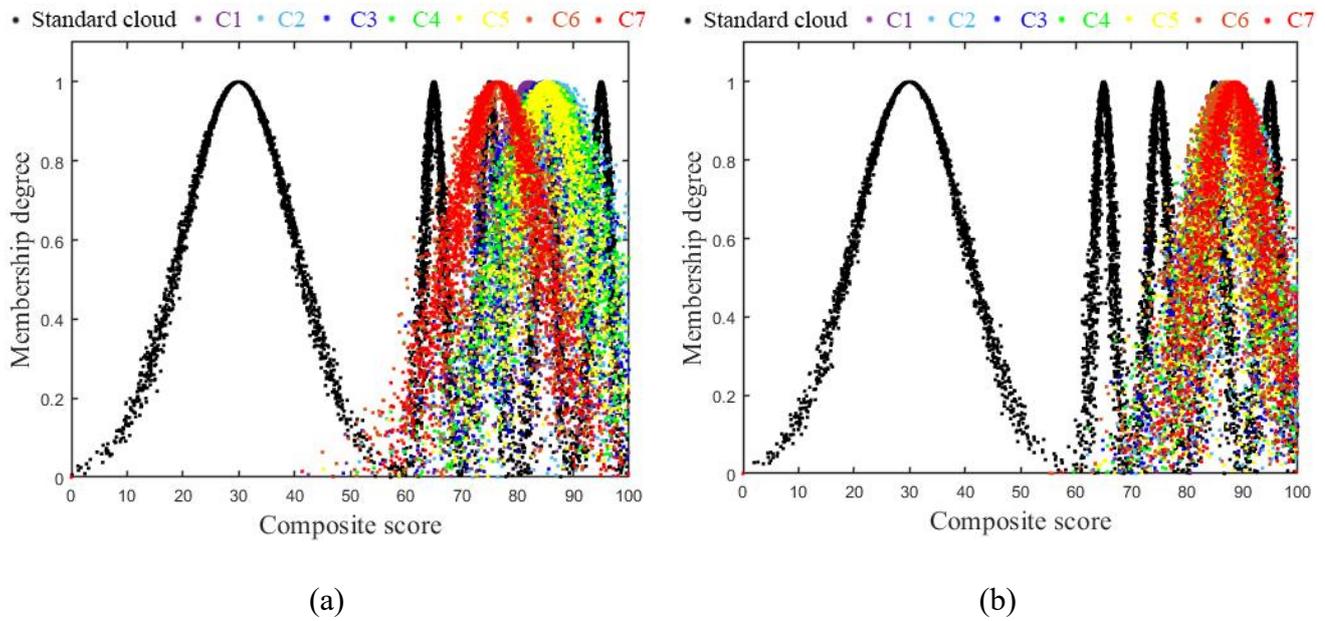

**Fig. 8.** Secondary-level indicators cloud diagram: (a) before planning, (b) after planning.

This analysis not only provides empirical foundation for deepening Tianmen's energy revolution implementation, but also establishes a replicable planning framework for global rural low-carbon transitions, thereby advancing the synergistic realization of energy security, economic development, and ecological conservation through integrated system optimization.

## 4. Conclusion

In this study, the evaluation index system of the new rural integrated energy system and the evaluation model based on the AHP-EWM and cloud model are constructed. Through a case analysis of City's rural energy revolution pilot and comparative model validation, the methodological soundness and real-world applicability of the proposed framework are confirmed. The weights determined by the AHP-EWM combined with the sum of deviation squares are reasonable and reliable, taking into account both subjective experience and objective data. The Cloud Model effectively addresses uncertainties in the evaluation process, enhancing result accuracy and robustness.

Through case analysis of Tianmen City's rural integrated energy system, this study demonstrates that planning of Tianmen City's could achieve triple enhancements: comprehensive evaluation (Ex increased from 83.12 to 87.55), operational stability (En decreased from 6.931 to 5.336), and risk controllability (He reduced from 3.08 to 2.278). The analysis reveals a 5.86% energy efficiency gain driven by multi-energy complementarity technologies, where intelligent regulation and optimized energy storage effectively mitigated renewable energy output fluctuations. Concurrently, social benefits surged by 14.1%, directly linked to improved power supply reliability and expanded employment in renewable sectors. Furthermore, environmental impact control has been strengthened by 29.8%, attributable to enhanced predictive capabilities in pollutant emission management. The research results provide important theoretical support and practical guidance for the optimal planning and scientific management of the new rural integrated energy system.

**Acknowledgements:**

The authors sincerely thank the engineers from Power China Hubei Engineering Co., Ltd. for their technical insights on rural energy system deployment.

**Reference:**
[1] Yin S, Zhao Y, Hussain A, Ullah K. Comprehensive evaluation of rural regional integrated clean energy systems considering multi-subject interest coordination with pythagorean fuzzy information. Engineering Applications of Artificial Intelligence 2024;138:109342. https://doi.org/10.1016/j.engappai.2024.109342.
[2] Mancarella P. MES (multi-energy systems): An overview of concepts and evaluation models. Energy 2014;65:1–17. https://doi.org/10.1016/j.energy.2013.10.041.
[3] Zhang J, Gao C, Wang T, Duan Y, Xu M, Guo Z. A Dynamic Evaluation Method for High-Permeability New Energy Distribution Network Planning Considering Multistage Development Trends. Front Energy Res 2022;10:958892. https://doi.org/10.3389/fenrg.2022.958892.
[4] Zhou J, Wu Y, Wu C, Deng Z, Xu C, Hu Y. A hybrid fuzzy multi-criteria decision-making approach for performance analysis and evaluation of park-level integrated energy system. Energy Conversion and Management 2019;201:112134. https://doi.org/10.1016/j.enconman.2019.112134.

[5] Li Y, Liu C, Zhang L, Sun B. A partition optimization design method for a regional integrated energy system based on a clustering algorithm. Energy 2021;219:119562. https://doi.org/10.1016/j.energy.2020.119562.

[6] Sheng K, Li Y, Li J, Chen Y, Zou J, Zhang Y, et al. A Survey on Post-Evaluation Indicator System for Multi-Energy Infrastructure Investments. IEEE Access 2020;8:158875–82. https://doi.org/10.1109/ACCESS.2020.3020548.

[7] Peng X, Guan X, Zeng Y, Zhang J. Artificial Intelligence-Driven Multi-Energy Optimization: Promoting Green Transition of Rural Energy Planning and Sustainable Energy Economy. Sustainability 2024;16:4111. https://doi.org/10.3390/su16104111.

[8] He Y, Zhang Y, Pang Y. Comprehensive Evaluation of Global Clean Energy Development Level Based on Compatibility and Difference Degree. Mathematical Problems in Engineering 2019;2019:6729209. https://doi.org/10.1155/2019/6729209.

[9] Chen X, Chen C, Tian G, Yang Y, Zhao Y. Comprehensive evaluation research of hybrid energy systems driven by renewable energy based on fuzzy multi-criteria decision-making. Front Energy Res 2023;11:1294391. https://doi.org/10.3389/fenrg.2023.1294391.

[10] Yang P, Jiang H, Liu C, Kang L, Wang C. Coordinated optimization scheduling operation of integrated energy system considering demand response and carbon trading mechanism. International Journal of Electrical Power & Energy Systems 2023;147:108902. https://doi.org/10.1016/j.ijepes.2022.108902.

[11] Zhao Y, Lv X, Shen X, Wang G, Li Z, Yu P, et al. Determination of Weights for the Integrated Energy System Assessment Index with Electrical Energy Substitution in the Dual Carbon Context. Energies 2023;16:2039. https://doi.org/10.3390/en16042039.

[12] Tan J, Pan W, Li Y, Hu H, Zhang C. Energy-sharing operation strategy of multi-district integrated energy systems considering carbon and renewable energy certificate trading. Applied Energy 2023;339:120835. https://doi.org/10.1016/j.apenergy.2023.120835.

[13] Zhao Z, Li H, Wang X, Li Z, Yu Z, Zhang Z, et al. Research on Evaluation Index System of Urban Energy Internet Development. IOP Conf Ser: Earth Environ Sci 2020;446:022052. https://doi.org/10.1088/1755-1315/446/2/022052.

[14] Nie J, Xu N, Ling Y, Xu N. Research on the Mixed Multi-attribute Group Evaluation System of Integral Energy System of Industrial Park Based on SNA Analysis Method. IOP Conf Ser: Earth Environ Sci 2020;526:012105. https://doi.org/10.1088/1755-1315/526/1/012105.

[15] Zhao Q, Du Y, Zhang T, Zhang W. Resilience index system and comprehensive assessment method for distribution network considering multi-energy coordination. International Journal of Electrical Power & Energy Systems 2021;133:107211. https://doi.org/10.1016/j.ijepes.2021.107211.

[16] Elbasuony GS, Abdel Aleem SHE, Ibrahim AM, Sharaf AM. A unified index for power quality evaluation in distributed generation systems. Energy 2018;149:607–22. https://doi.org/10.1016/j.energy.2018.02.088.

[17] Leng Y-J, Zhang H. Comprehensive evaluation of renewable energy development level based on game theory and TOPSIS. Computers & Industrial Engineering 2023;175:108873. https://doi.org/10.1016/j.cie.2022.108873.

[18] Li J, Xu W, Feng X, Lu H, Qiao B, Gu W, et al. Comprehensive evaluation system for optimal configuration of multi-energy systems. Energy and Buildings 2021;252:111367. https://doi.org/10.1016/j.enbuild.2021.111367.

[19] Fan L, Zhong X, Chen J, Tu Y, Jiang J. Comprehensive Energy System Evaluation Method of Low-carbon Park Based on Improved Entropy Weight. J Phys: Conf Ser 2023;2418:012025. https://doi.org/10.1088/1742-6596/2418/1/012025.

[20] Li Z, Wang Y, Xie J, Cheng Y, Shi L. Hybrid multi-criteria decision-making evaluation of multiple renewable energy systems considering the hysteresis band principle. International Journal of Hydrogen Energy 2024;49:450–62. https://doi.org/10.1016/j.ijhydene.2023.09.059.

[21] Wang W, Li H, Hou X, Zhang Q, Tian S. Multi-Criteria Evaluation of Distributed Energy System Based on Order Relation-Anti-Entropy Weight Method. Energies 2021;14:246. https://doi.org/10.3390/en14010246.

[22] Shi B, Wang H. Policy effectiveness and environmental policy Assessment: A model of the environmental benefits of renewable energy for sustainable development. Sustainable Energy Technologies and Assessments 2023;57:103153. https://doi.org/10.1016/j.seta.2023.103153.

[23] Wang Y, Shi L, Song M, Jia M, Li B. Evaluating the energy-exergy-economy-environment performance of the biomass-photovoltaic-hydrogen integrated energy system based on hybrid multi-criterion decision-making model. Renewable Energy 2024;224:120220. https://doi.org/10.1016/j.renene.2024.120220.

[24] Ke Y, Liu J, Meng J, Fang S, Zhuang S. Comprehensive evaluation for plan selection of urban integrated energy systems: A novel multi-criteria decision-making framework. Sustainable Cities and Society 2022;81:103837. https://doi.org/10.1016/j.scs.2022.103837.

[25] Kong M, Ye X, Liu D, Li C. Comprehensive evaluation of medical waste gasification low-carbon multi-generation system based on AHP–EWM–GFCE method. Energy 2024;296:131161. https://doi.org/10.1016/j.energy.2024.131161.

[26] Zhao H, Li B, Lu H, Wang X, Li H, Guo S, et al. Economy-environment-energy performance evaluation of CCHP microgrid system: A hybrid multi-criteria decision-making method. Energy 2022;240:122830. https://doi.org/10.1016/j.energy.2021.122830.

[27] Yang K, Ding Y, Zhu N, Yang F, Wang Q. Multi-criteria integrated evaluation of distributed energy system for community energy planning based on improved grey incidence approach: A case study in Tianjin. Applied Energy 2018;229:352–63. https://doi.org/10.1016/j.apenergy.2018.08.016.

[28] Yang Z, Wang X. Research on Low-Carbon Capability Evaluation Model of City Regional Integrated Energy System under Energy Market Environment. Processes 2022;10:1906. https://doi.org/10.3390/pr10101906.

[29] Xuan K, Hao Y, Liang Z, Zhang J. Research on the evaluation of distributed integrated energy system using improved analytic hierarchy process-information entropy method. Energy Sources, Part A: Recovery, Utilization, and Environmental Effects 2022;44:10071–93. https://doi.org/10.1080/15567036.2022.2143951.

[30] Zhao H, Guo S. Urban integrated energy system construction plan selection: a hybrid multi-criteria decision-making framework. Environ Dev Sustain 2024. https://doi.org/10.1007/s10668-024-04491-y.

[31] Zhang Y, Wang R, Huang P, Wang X, Wang S. Risk evaluation of large-scale seawater desalination projects based on an integrated fuzzy comprehensive evaluation and analytic hierarchy process method. Desalination 2020;478:114286. https://doi.org/10.1016/j.desal.2019.114286.

[32] Li Y, Chen Y, Li Q. Assessment analysis of green development level based on S-type cloud model of Beijing-Tianjin-Hebei, China. Renewable and Sustainable Energy Reviews 2020;133:110245. https://doi.org/10.1016/j.rser.2020.110245.

[33] Qin G, Zhang M, Yan Q, Xu C, Kammen DM. Comprehensive evaluation of regional energy internet using a fuzzy analytic hierarchy process based on cloud model: A case in China. Energy 2021;228:120569. https://doi.org/10.1016/j.energy.2021.120569.

[34] Liu W, Zhu J. A multistage decision-making method for multi-source information with Shapley optimization based on normal cloud models. Applied Soft Computing 2021;111:107716. https://doi.org/10.1016/j.asoc.2021.107716.

[35] Wu Z, Yang C, Zheng R. Developing a holistic fuzzy hierarchy-cloud assessment model for the connection risk of renewable energy microgrid. Energy 2022;245:123235. https://doi.org/10.1016/j.energy.2022.123235.

[36] Nie S, Cai G T, Gao L P. A 3E (Energy, Economy, Environment) Evaluation Method for Park-Scale Integrated Energy Systems Based on Game Theory-Cloud Model. *Advances in New Energy* 2022;10:169–177. https://doi.org/10.3969/j.issn.2095-560X.2022.02.010. (in Chinese)

[37] Berjawi AEH, Walker SL, Patsios C, Hosseini SHR. An evaluation framework for future integrated energy systems: A whole energy systems approach. Renewable and Sustainable Energy Reviews 2021;145:111163. https://doi.org/10.1016/j.rser.2021.111163.

[38] Wang Y, Zhang L, Song Y, Han K, Zhang Y, Zhu Y, et al. State-of-the-art review on evaluation indicators of integrated intelligent energy from different perspectives. Renewable and Sustainable Energy Reviews 2024;189:113835. https://doi.org/10.1016/j.rser.2023.113835.

[39] Mo L, Liu X, Chen H, Zhao Z, Chen J, Deng Z. Exergy-economic analysis and evaluation method of park-level integrated energy system. Front Energy Res 2022;10. https://doi.org/10.3389/fenrg.2022.968102.

[40] Luo XJ, Oyedele LO, Owolabi HA, Bilal M, Ajayi AO, Akinade OO. Life cycle assessment approach for renewable multi-energy system: A comprehensive analysis. Energy Conversion and Management 2020;224:113354. https://doi.org/10.1016/j.enconman.2020.113354.

[41] Liu H, Shen W, He X, Zeng B, Liu Y, Zeng M, et al. Multi-scenario comprehensive benefit evaluation model of a multi-energy micro-grid based on the matter-element extension model. Energy Science & Engineering 2021;9:402–16. https://doi.org/10.1002/ese3.828.

[42] Zhao D, Li C, Wang Q, Yuan J. Comprehensive evaluation of national electric power development based on cloud model and entropy method and TOPSIS: A case study in 11 countries. Journal of Cleaner Production 2020;277:123190. https://doi.org/10.1016/j.jclepro.2020.123190.

[43] Xu Z, Liao H. Intuitionistic Fuzzy Analytic Hierarchy Process. IEEE Trans Fuzzy Syst 2014;22:749–61. https://doi.org/10.1109/TFUZZ.2013.2272585.

[44] Li G, Li J P, Sun X L, Zhao M. A Study on the Combination Methods of Subjective and Objective Weights and Their Rationality. *Management Review* 2017;29:17–26, 61. https://doi.org/10.14120/j.cnki.cn11-5057/f.2017.12.002. (in Chinese)

[45] Gong Y B, Xu B X, Liu G F. Comprehensive Similarity Measure of Cloud Model and Its Application in Linguistic Multi-Attribute Decision Making. Journal of Systems Science and Mathematical Sciences 2024;44:3371–87. https://doi.org/12341/jssms23357. (in Chinese)

[46] Tang M, Li R, Zhang R, Yang S. Research on New Electric Power System Risk Assessment Based on Cloud Model. Sustainability 2024;16:2014. https://doi.org/10.3390/su16052014.

[47] Deng Z, Du A, Yang C, Tong J, Chen Y. Multilevel Evaluation Model of Electric Power Steering System Based on Improved Combination Weighting and Cloud Theory. Applied Sciences 2024;14:1043. https://doi.org/10.3390/app14031043.

[48] Wu K, Wang Y, Qi L, Liu R, Liu D, Zhang Y. Operation Mode and Value Evaluation System of Integrated Energy System in the Park. IOP Conf Ser: Earth Environ Sci 2020;526:012129. https://doi.org/10.1088/1755-1315/526/1/012129.